\documentclass[aps,prl,twocolumn,showpacs,showkeys,citeautoscript]{revtex4-1}

\usepackage{graphicx}
\usepackage{amssymb}
\usepackage{amsmath}

\newcommand{\ud}{\,{\mathrm d}}

\newcommand{\tc}{\overline{t}}

\newcommand{\uiiint}{\int\!\!\!\int\!\!\!\int}

\begin{document}

\title{Large vortex state in ferromagnetic disks.}

\author{Konstantin L. Metlov$^1$}

\affiliation{Donetsk Institute of Physics and Technology NAS, Donetsk, Ukraine 83114}
\email{metlov@fti.dn.ua}
\date{\today}
\begin{abstract}
Magnetic vortices in soft ferromagnetic nano-disks have been
extensively studied for at least several decades both for their
fundamental (as a ``live'' macroscopic realization of a field theory
model of an elementary particle) as well as applied value for
high-speed high-density power-independent information storage. Here it
is shown that there is another vortex state in nano-scale
ferromagnetic disks of several exchange lengths in size. The energy of
this large vortex state is computed numerically (within the framework
of Magnetism@home distributed computing project) and its stability is
studied analytically, which allows to plot it on magnetic phase
diagram. It is the ground state of cylinders of certain sizes and is
metastable in a wider set of geometries. Large vortices exist on par
with classical ones, while being separated by an energy barrier,
controllable by tuning the geometry and material of ferromagnetic
disk. This state can be an excellent candidate for magnetic
information storage not only because the resulting disk sizes are
among the smallest, able to support magnetic vortices, but also
because it is the closest to the classical vortex state of all other
known metastable states of magnetic nano-cylinder, which implies, that
the memory, based on switching between these two different types of
magnetic vortices, may, potentially, achieve the highest possible rate
of switching.
\end{abstract}
\maketitle

Information storage is always based on switching some physical system
between different metastable states, separated by energy barrier. In
the case of magnetic nano-cylinders with magnetic vortices
(theoretically studied by N.A. Usov and S.E. Peschany\cite{UP93} and
confirmed experimentally by Shinjo et al\cite{SOHSO00}), the vortex
states of different vortex core polarizations\cite{VanWaeyenberge2006}
or different chiralities\cite{AO2009} are now being extensively
researched for this purpose. There is much progress in this
area\cite{Kammerer2011}, including the remarkable results on switching
the vortex core polarity by ultra-short in-plane magnetic field
pulses\cite{HGFS07,LGLK07,GK08}. However, despite the driving pulses
can be made extremely short, the switching itself is usually
accompanied by much longer magnetization dynamics during which the new
equilibrium state is established\cite{SCSFM03}. If the switching
process involves creation of additional vortices and anti-vortices,
they must be annihilated at the end, which implies significant
spin-wave generation\cite{CKGK07}. The energy of these spin-waves (as
well as other energy, accumulated by magnetization) must be
dissipated, which limits the sustained {\em rate of switching} of such
a devices. This limitation is more pronounced, the longer is the
trajectory, each spin must pass during the switching, or, the bigger
is the distance between the metastable states used. Knowing these
states is, therefore, very important for applications and
also is an intriguing fundamental problem. Similarity of the
equations means that finding a new metastable state in nano-magnetism
is like discovering a new elementary particle in non-linear field
theory. With a difference that nano-magnets have boundary, allowing
for existence of additional states, such as boundary-bound half-
vortices and anti-vortices\cite{M10} as well as large vortices.

Because the equations for equilibrium magnetic textures are non-linear
and non-local, solving them for stationary states usually involves
significant guesswork. Usually, new states (such as domain structures
or new types of domain walls) are first observed in experiments (or
numerical experiments) and only then described theoretically. The
experiment also does not provide researchers with a ``silver bullet'',
since there are many metastable states in magnetism and discovering a
new one experimentally (often by accident) involves applying a certain
(sometimes quite complex) sequence of external forces to the system
(or, setting initial conditions in the case of numerical experiments),
which (may or may not) drive it to a new stable state. This task
becomes especially hard if the new state occupies only a small part of
magnetic phase diagram or appears only as metastable (as opposed to
the lowest energy ground state of the system). In case of
nano-magnets, however, it is possible to parametrize the low-energy
magnetization distributions using several scalar parameters\cite{M10},
which facilitates systematic exploration of their state space. Such
exploration of states in circular cylinder was performed in the
framework of distributed computing project Magnetism@home.

Specifically, the project is based on the following parametrization of
magnetic texture:
\begin{equation}
\label{eq:ft}
f^{M@H}(t)=c \left(\frac{\imath t}{p}-\frac{1}{2}\left(a - 
\frac{\overline{a}t^2}{p^2}\right)\right) T'(t),
\end{equation}
which specifies components of dimensionless magnetization vector
$\vec{m}(t)=\vec{M}/M_S$, normalized by saturation magnetization of
cylinder's material $M_S$, via stereographic projection $m_x+\imath
m_y=2 f/(1+|f|^2)$, $m_z=(1-|f|^2)/(1+|f|^2)$. The texture is assumed
to be uniform along the cylinder axis ($Z$) and depends only on the
in-plane complex coordinate $t=X+\imath Y$, where $X$, $Y$ and $Z$ are
dimensionless Cartesian coordinates, normalized by the cylinder radius
$R$ so that $|t|=X^2+Y^2 \le 1$.  Along $Z$ axis the cylinder is located
at $-g/2\le Z \le g/2$, where $g=L_Z/R$ is normalized
thickness (aspect ratio).

The texture (\ref{eq:ft}) is a particular single-vortex case of a more
general texture\cite{M10} with additional rescaling, governed by the
parameter $p>1$, to cover states with quasi-uniform
magnetization\cite{MG04}. It can be applied to cylinders of arbitrary
shape of their face, which is specified by the conformal map $T(t)$
from unit disk $|t|\le 1$ to the desired shape. In this paper (and in
the first run of the Magnetism@home project) only the circular
cylinders are considered, that is $T(t)=t$ and $T'(t)=1$. Complex
phase of parameter $a$ controls direction of the vortex center
displacement. In circular cylinder this phase can be nullified by
rotating the coordinate system. The parameter $a$ was, therefore,
assumed to be real $0 \le a<\infty$. The parameter $0<c<\infty$
controls the size of the vortex core (or half-vortex cores, if they
are present).

Magnetism@home project numerically computed magnetostatic energy
(including that of volume, face and side magnetic charges, as well as
their mutual interaction) of magnetization distributions (\ref{eq:ft})
in a 4-dimensional unit-hypercube with normalized coordinates
$\widetilde{g}=g/(1+g)$, $\widetilde{c}=c/(1+c)$,
$\widetilde{a}=a/(1+a)$, $\widetilde{p}=1/p$, which covers all the
magnetization configurations it describes. The densities of magnetic
charges were computed analytically, while the magnetostatic energy
itself was evaluated using fast multipole method\cite{VA2003FMM} on a
dense 50000 non-uniform finite elements mesh, perfectly covering the
circular cylinder's boundary, with exact analytical treatment for $Z$
dependence of demagnetizing field. This computation (including
preliminary testing runs) took about a year to complete, using idle
processor time of a several tens of thousands of computers on the
Internet, communicating via Berkeley Open Infrastructure for Network
Computing (BOINC) protocol. The exchange energy is much simpler to
evaluate (and also its dependence on $g$ can be taken into account
analytically), which was done locally. The resulting data files
(including the source code of programs, used to compute and process
them) are published as Magnetism@home data release 1, attached as
supplemental material.

Combining and interpolating Magnetism@home files allows to compute
multi-dimensional energy landscape of the particle as function of $p$,
$a$ and $c$ in particles of different physical dimensions $R/L_E$ and
$L_Z/L_E$, where $L_E=\sqrt{C/\gamma_B \mu_0 M_S}$ is the exchange
length of cylinder's material with exchange stiffness $C$, $\mu_0$ is
permeability of vacuum and $\gamma_B=1$ (in CGS system of
units\cite{AharoniBook} $\mu_0=1$ and $\gamma_B=4\pi$). Then, building
the ground state magnetic phase diagram is as simple as finding the
smallest value in the resulting energy array at each point
($R/L_E$,$L_Z/L_E$) and classifying the corresponding state (given by
values of $p$, $a$ and $c$ at minimum). The resulting numerical
diagram is shown in Figure by shading.
\begin{figure*}
\label{fig:diagram}
 \includegraphics[scale=0.55]{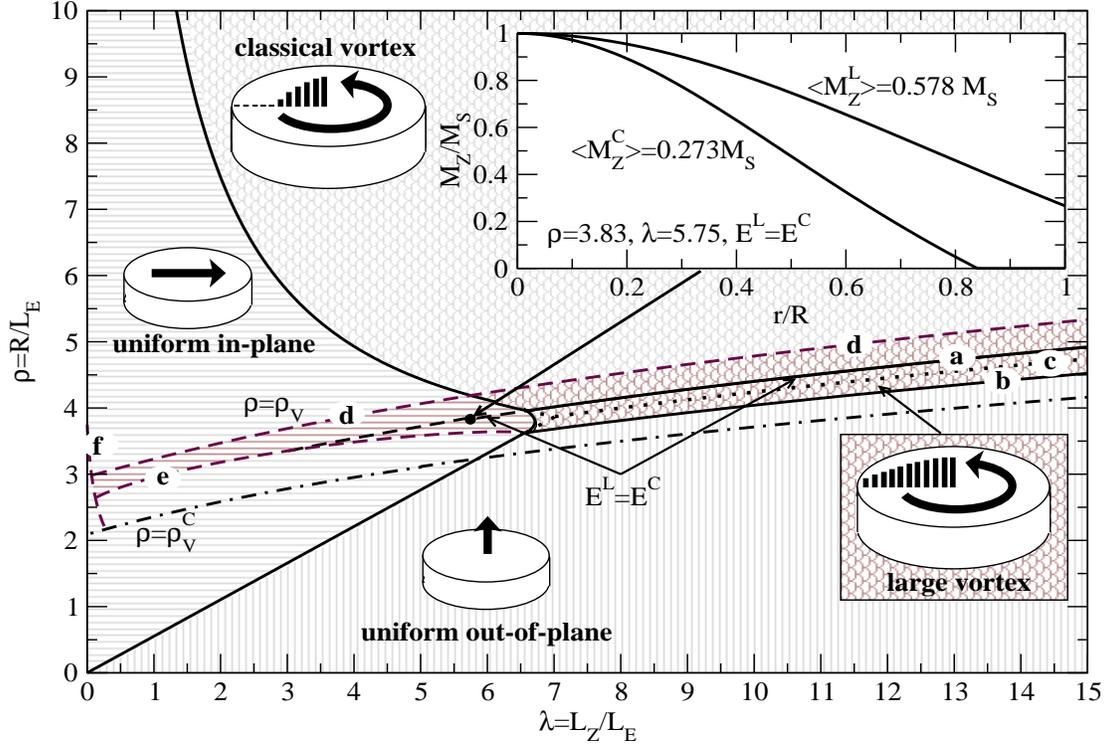}
\caption{Equilibrium and stability of large magnetic vortex state in
  circular cylinders of different radii $R$ and thicknesses $L_Z$ in
  soft magnetic material with exchange length $L_E$ and saturation
  magnetization $M_S$. Solid lines separate the regions (also having
  different texture), where different states (uniform in-plane,
  uniform out-of-plane, classical vortex and large vortex) have the
  lowest energy. They correspond to particle geometries, where the
  energy of respective pairs of states is equal. Cylinder geometries,
  where large magnetic vortex can be metastable, are shown by darker
  shading. Lettered lines, outlining this stability region, are
  discussed in the text, they correspond to different modes of vortex
  stability loss. The line $\rho_V^C$ shows the radius of classical
  vortex (which depends only on cylinder's thickness), computed by
  Usov and Peschany\cite{UP93}. Inset shows out-of-plane magnetization
  component distribution $M_Z/M_S$ as function of reduced radial
  coordinate $r/R$ in large and classical vortex at a particular
  geometry, where they both are stable and have the same energy
  ($E^L=E^C$).}
\end{figure*}

Potentially, the files allow to extract much more information, but for
now, let us focus on the ground state diagram, as it already displays
a new simple and fundamental result: there is another magnetic vortex
state, which, in tall cylinders ($L_Z \gtrsim 6.5$) of certain radii,
is, actually, the ground state. At larger cylinder radii the energy of
this state becomes equal (shown by the solid line {\it a} in Figure)
and then larger than the energy of the classical magnetic vortex. At
smaller radii it continuously transforms into the state of completely
uniform out-of-plane magnetization (solid line {\it b}). This expands
the region of existence of magnetic vortices to particles of smaller
radii and solves a well known paradox\cite{GN04_JAP_PERP,ML08}
that (when the large vortices are not taken into account) the uniform
out-of-plane state continues to be the ground state of the cylinder
(with energy, smaller than that of classical vortex) with geometries
(below the dotted line {\it c} and above line {\it b}), where it is no
longer stable with respect to transformation into magnetic vortex. The
line {\it b} corresponds to the second order phase transition during
which the uniform state loses stability and total Z-component of
magnetization starts to deviate from $M_S$. The line {\it a}
corresponds to the first order transition, accompanied by hysteresis,
so that the large vortices may exist as metastable in certain region
above it, while the classical vortices may exist below this line (they
continue to be stable with respect to core expansion in cylinders with
radii down to equilibrium classical vortex core radius, shown in
Figure as dash-dotted line $\rho_V^C$).

To confirm the above numerical results and also to fully outline the
region of stability of the large magnetic vortices let us now proceed
with analytical consideration.

Two interactions, present in any ferromagnet, are the exchange and
magnetostatic self-interaction. In continuum approximation, assuming
the magnetization texture is simply a function of coordinates
$\vec{m}(\vec{r})$, $r=\{X,Y,Z\}$, the exchange energy can be
represented as
\begin{equation}
E_{EX}=\frac{C M_S}{2} \uiiint_V 
\sum\limits_{i=X,Y,Z}(\vec{\nabla} m_i(\vec{r}))^2 \ud^3\vec{r},
\end{equation}
where integration is performed across the volume of the magnet $V$,
outside of the magnet $|m|=0$. In the considered case of vortices,
larger then the particle radius, the magnetization texture can be
described by the following analytical function of complex variable
\begin{equation}
  \label{eq:fcl}
  f(t)=\frac{\imath t}{\rho_V},
\end{equation}
with $\rho_V=R_V/R>1$ is the normalized vortex core radius $R_V$. Since
$\rho_V>1$, this magnetization texture is all-soliton\cite{M10}. That
is, contrary to the case of classical magnetic vortex, there is no
region in the particle, where magnetization vector lies in the
cylinder plane. In this case, the exchange energy can be expressed
directly in terms of $f(z)$ as
\begin{equation}
  \label{eq:ex}
  E_{EX}=4 C M_S \uiiint_V 
  \frac{|f'(t)|^2}{(1+|f'(t)|^2)^2}
  \ud^3\vec{r},
\end{equation}
which, after integration, gives
\begin{equation}
  e_{EX}^L=\frac{4}{\rho^2(1+\rho_V^2)},
\end{equation}
where the dimensionless energy $e=E/(\mu_0\gamma_B M_S^2 \pi L_Z R^2)$
and dimensionless particle radius $\rho=R/L_E$ were introduced.

To compute the magnetostatic energy let us employ magnetic charges
formalism, following from Maxwell's equations in the static case
with no macroscopic currents present. Then, it is possible to
express the demagnetizing field via gradient of a scalar potential,
which is solution of Poisson's equations with the volume density of
magnetic charge $\Omega=-\mathrm{div}\vec{M}$ on the right hand side. On the
boundary of magnetic material this volume density reduces to localized
surface density of magnetic charge, equal to the component of
magnetization vector, normal to the boundary. In the case of centered
large vortex (\ref{eq:fcl}) only the surface magnetic charge on the
particle face, proportional to out-of-plane component of magnetization
$\sigma=M_z$ is present. Its energy can be expressed as
\begin{equation}
  e_{MS}=\frac{1}{g} \left(w\left(\rho_V,0\right)- w\left(\rho_V,g\right)\right),
\end{equation}
where the magnetostatic function
\begin{equation}
 w(\rho_v,g)=\!\!\int\limits_0^1 \!
\int\limits_0^{2\pi} \! \int\limits_0^1 \!
\int\limits_0^{2\pi} \!\!
\frac{\sigma(r_1,\varphi_1)\sigma(r_2,\varphi_2) 
  r_1\ud r_1\ud\varphi_1 r_2\ud r_2\ud\varphi_2}
     {\sqrt{r_1^2\!+\!r_2^2\!-
         \!2 r_1 r_2 \cos (\varphi_1\!-\!\varphi_2)\! +\! g^2}},
\end{equation}
with $r$ and $\varphi$ being dimensionless coordinates on cylinder's
face. For centered vortex (\ref{eq:fcl}) the density of face
charges does not depend on $\varphi$ and the integral over the angles can
be factored out, producing
\begin{eqnarray}
  & & w(\rho_v,g)=  \\
  & & \frac{2}{\pi}
  \int\limits_0^1\!\int\limits_0^1
  \frac{r_1 r_2 (\rho_V^2 - r_1^2) (\rho_V^2 - r_2^2) 
    K(-\frac{4 r_1 r_2}{g^2+(r1-r2)^2})\ud r_1 \ud r_2}
  {(\rho_V^2 + r_1^2) (\rho_V^2 + r_2^2)
  \sqrt{g^2+(r1-r2)^2}},
\end{eqnarray}
where $K(k)$ is complete elliptic integral of the first kind. The
function $w$ is a nice continuous function and can be differentiated
over $\rho_v$ under the integral.

To find equilibrium radius of large magnetic vortex it is now
sufficient to differentiate the total energy $e_{EX}+e_{MS}$ over
$\rho_V$ and require that this derivative is equal to 0, which leads
to the following transcendental equation
\begin{equation}
  -\frac{8 g \rho_v}{\rho^2(1+\rho_V^2)^2} +
  \frac{\partial}{\partial \rho_V} \left( 
  w\left(\rho_V,0\right)- w\left(\rho_V,g\right)
  \right) = 0.
\end{equation}
This equation is hard to solve for $\rho_V$. However, it is easy to
solve for $\rho$, which gives
\begin{equation}
  \label{eq:rrv}
  \rho = \frac{2}{1+\rho_V^2}\sqrt{
    \frac{2 g \rho_V}{\partial_{\rho_V}w\left(\rho_V,0\right)- \partial_{\rho_V}w\left(\rho_V,g\right)}
  },
\end{equation}
allowing to compute the radius of the cylinder $\rho$ in which the
large vortex of specified radius $\rho_V$ would be at equilibrium. The
limit $\rho_V \rightarrow \infty$ of this expression immediately
recovers the line {\it b} in Figure -- the boundary at which the large
vortices (at larger radii) lose their chirality and
become the uniform out-of-plane state (at smaller radii).

The other stability boundary of the large vortex state, line {\it d},
can be found in two different ways. One is direct but cumbersome, by
considering the transformation between the all-soliton magnetic vortex
and the classical one, looking for the particle geometry where the
former becomes unstable with respect to transformation into the
latter. This can be achieved by considering magnetization
distributions, expressed via the complex function
\begin{equation}
  w(t,\tc)=\left\{
    \begin{array}{ll}
      f(z) & |f(z)| \leq 1 \\
      \frac{f(t)}{|f(t)| \sin^2 \alpha + \cos^2\alpha} & |f(z)| > 1
    \end{array}
\right.
 .
\end{equation}
This function is, in general, not analytical. At $\alpha=0$ it
describes all-soliton vortex and at $\alpha=\pi/2$ non-analytical
everywhere classical vortex, having a meron part at $|t|>\rho_V$. The
line {\it d} can be obtained as a boundary at which the energy minimum
at $\alpha=0$ disappears. Such analysis was indeed performed, however,
it turns out (at least to numerical precision) that all-soliton
vortices lose stability right at the moment their core boundary touches
the particle boundary. Then they immediately transform into classical
vortices of much smaller core radius $\rho_V^C$. Thus, the line {\it d} can be
found simply by computing the particle radius (\ref{eq:rrv}) in the
limit $\rho_V \rightarrow 1$.

There is another catch that the line {\it b} we have just computed
becomes unphysical when the aspect ratio of the particle $g=L_Z/R$
becomes smaller then approximately $1.81295$. At thinner cylinders the
uniform {\em in-plane} state becomes the ground state and it becomes
necessary to study the transformation of the large vortex state into
it. This can be done by considering the uniform tilt of the
magnetization in the large vortex, described by the complex function
\begin{equation}
  \label{eq:ftilt}
  f^{\mathrm{tilt}}(t)=\frac{\imath f(t) \cos (\alpha/2)+\sin(\alpha/2)}
  {\imath \cos (\alpha/2)+f(t) \sin(\alpha/2)}.
\end{equation}
At $\alpha=0$ it is a pure large vortex (\ref{eq:fcl}), but, as
$\alpha$ increases, it's magnetization uniformly rotates towards the
particle plane. The stability line, then, corresponds to the
disappearance of the energy minimum at $\alpha=0$ and can be found by
solving the equation
\begin{equation}
  \label{eq:tilt}
\left.\frac{\partial^2 e^{\mathrm{tilt}}}{\partial \alpha^2}\right|_{\alpha=0}=\left.\frac{\partial^2 e_{EX}^{\mathrm{tilt}}}{\partial \alpha^2}+\frac{\partial^2 e_{MS}^{\mathrm{tilt}}}{\partial \alpha^2}\right|_{\alpha=0}=0
\end{equation}
at equilibrium $\rho_V$. Since the $f^{\mathrm{tilt}}(t)$ is still an
analytical function of $t$ the exchange energy
$e_{EX}^{\mathrm{tilt}}$ can be directly obtained from
(\ref{eq:ex}). The magnetostatic energy, however, now has volume and
side surface charge contribution in addition to the energy of face
charges. Moreover, there is an interaction between the side and volume
charges (their interaction with face charges is canceled by
symmetry), which also depend on the polar angle $\varphi$. This makes
it necessary to adopt another method of computing magnetostatic energy
as angular integrals can't be directly factored-out. First, let us
merge the volume and side charges density together, using Dirac's
delta function $\delta(x)$ to express the side charges
localization. Their product (entering the magnetostatic integral) is
\begin{equation}
  \left.\frac{\partial^2}{\partial \alpha^2}
  \Omega(r_1,\varphi_1,\alpha)  \Omega(r_2,\varphi_2,\alpha)
  \right|_{\alpha=0} =
  2\, \omega(r_1,\varphi_1) \omega(r_2,\varphi_2),
\end{equation}
where
\begin{equation}
  \omega(r,\varphi) =
  -\frac{4 r \rho_v^2 \sin \varphi}{(r^2 + \rho_V^2)^2}+
  \frac{(1-\rho_V^2)\delta(1-r)\sin \varphi}{1+\rho_V^2}.
\end{equation}
The integral of such product of charges can be very efficiently
evaluated\cite{GM01} by representing the inverse distance in polar
coordinates via Lipshitz internal and then using the Bessel's
summation theorem to factor the angular integral out, which gives
\begin{equation}
\left.\frac{\partial^2 e_{MS}^{\mathrm{tilt}}}{\partial \alpha^2}\right|_{\alpha=0}=
\frac{1}{2}\int_0^\infty \frac{f_{MS}(k g)}{k}
\left(\int_0^1
\omega(r,\varphi) J_1(k r) r
\ud r
\right)^2 \ud k,
\end{equation}
where $J_1(x)$ is Bessel's function of the first kind and the delta
function is assumed to be right-sided. This integral was evaluated
numerically, which is sufficient to solve the transcendental equation
(\ref{eq:tilt}) and compute the stability line {\em e}. Below this
line the large vortices, if created, immediately convert into the
uniform in-plane state. A confirmation of the consistency of the
performed computations is the fact that the line {\it e} joins the line
{\it b} exactly at the critical aspect ratio $g=1.81295$ at which the
energies of uniform in-plane and out-of-plane states are equal.

Finally, there is another possibility of vortex conversion into the
uniform in-plane state, not captured by the expression
(\ref{eq:ftilt}). At very small thicknesses the vortex may become
unstable with respect to lateral shift. The stabilizing force in
this case is produced by the side magnetic charges, whose role quickly
diminishes as particle becomes thinner. The corresponding stability
line {\it f} can be found by equating to zero the second derivative in
$a$ at $a=0$ of the energy of magnetization distribution,
corresponding to the following complex function
\begin{equation}
  f^{\mathrm{shift}}(t)=\frac{\imath (z-a)}{\rho_V}.
\end{equation}
The face charges energy is unchanged in this case, and the balance of
forces is established between the exchange (\ref{eq:ex}) and
magnetostatic interaction of side charges. The exchange prevails for
radii below the stability line {\it f} and the particle becomes
uniformly magnetized in-plane.

Let us now discuss limitations of the above consideration. In reality
there will be a deviation from the assumption of magnetization
uniformity across cylinder's thickness. However, its influence on the
phase diagram lines (in the considered range of thicknesses) is
negligible, as evidenced by 3-d micromagnetic simulations and
experiment\cite{Scholz2003}. Also, the metastability region of large
vortex state comes down to $L_Z/L_E\sim0.3$, where magnetization
thickness independence is beyond any doubt. Full 3-d models can,
however, bring in new effects. At large thicknesses the opposite faces
of cylinder become magnetostatically decoupled, so in thicker cut
cones (such as commonly produced by lithography), it might be possible
to have a metastable large vortex state on smaller face, while having
only the small classical-like vortex stable on larger face. Such
states can be useful e.g. for readout of stored information, but their
existence needs a separate confirmation. Lithographic process may also
produce other defects, such as imperfect boundary or local variations
of thickness. While large deviations of this kind would produce
completely unpredictable magnetic configurations, small random defects
usually pin down states across the first order phase transition
boundary and make hysteresis loops wider. Experimental points in Fig
14 of Ref.~\onlinecite{Scholz2003} also indicate that dots with
dimensions, supporting large vortices, are experimentally accessible
and are not superparamagnetic. The exact lifetime analysis of these
states still needs to be performed, but from fundamental point of view
they are viable and must be kinetically stable at least at
sufficiently low temperatures.

Concluding, it was shown that there is another vortex state in
ferromagnetic circular cylinders with vortex core radius, larger than
the particle radius. In certain range of particle geometries this
large vortex may coexist with classical vortex\cite{UP93}, whose core
is fully inside the particle. In this region one may have the energies
of both classical vortex and large vortex equal (see inset in Figure)
while their magnetic moment differs significantly. It should be
possible to switch magnetization between these two vortex states
(e.g. by applying current through the particle). Moreover, because
these states can be continuously transformed into one another without
formation of Bloch points or other topological singularities and are
otherwise very similar to each other, such switching can be expected
to be very fast.

Acknowledgments. The distributed computation under the umbrella of
Magnetism@home project was made possible thanks to participation of
numerous volunteers, whose help is gratefully acknowledged.  I'd like
to thank Alexander E. Filippov for reading the manuscript and many
valuable suggestions.

%

\end{document}